\documentclass[twocolumn,preprintnumbers,amsmath,amssymb]{revtex4}
\usepackage{graphicx}% Include figure files
\usepackage{dcolumn}% Align table columns on decimal point
\usepackage{bm}% bold math
\begin{document}
\title{\bf Ground state study of simple atoms\\ within a nano-scale box}
\author{ M. Neek-Amal$^{1,2}$\thanks{Corresponding Author: e-mail: neek@nano.ipm.ac.ir,
Tel:(+98)212835058,Fax:(+98) 212280415.}, G. Tayebirad$^{3}$, M.
Molayem$^{3}$, \\M. E. Foulaadvand$^{1,3}$, L.
Esmaeili-Sereshki$^{5}$ and  A. Namiranian$^{1,3}$} \affiliation{
 $^1$Department of Nano-Science, Institute for Studies in Theoretical
 Physics\\ and Mathematics (IPM), P.O. Box 19395- 5531, Iran.\\
$^2$Department of Physics, Shahid Rajaei University, Lavizan, Tehran 16788, Iran.\\
 $^3$Department of Physics, Iran University of Science
and Technology, P.O. Box 16844, Iran.\\
 $^4$Department of Physics, Zanjan University, P.O. Box 45195-313, Zanjan,  Iran.\\
 $^5$Department of Physics, Alzahra University, P.O. Box 15815-3487, Tehran,  Iran.}
\begin{abstract}

Ground state energies for confined hydrogen (H) and helium (He)
atoms, inside a penetrable/impenetrable compartment have been
calculated using Diffusion Monte Carlo (DMC) method. Specifically,
we have investigated spherical and ellipsoidal encompassing
compartments of a few nanometer size. The potential is held fixed at
a constant value on the surface of the compartment and beyond.
The dependence of ground state energy on the geometrical characteristics of the
compartment as well as the potential value on its surface has been
thoroughly explored. In addition, we have investigated the cases
where the nucleus location is off the geometrical centre of the compartment.\\

\end{abstract}
\maketitle \vspace{1cm}

\newpage

\section{ Introduction }

In the past few years, spatially confined atoms and molecules have
gained relentless attention \cite{1}. This comes mainly from the
notable difference between the ground state properties of such
confined systems in comparison to their bulk states. As a result,
the confining systems may be applied in several circumstances. At
present, it is argued that physics of high pressure materials can be
investigated within models of confining particles. Furthermore,
extensive investigations over large classes of nano-structure
systems such as quantum dots and artificial atoms are essentially
related to the problem of confined atoms and electrons \cite{2}. By
the issue {\it confined atoms}, we mean the atoms that experience
external potentials which keep themselves in a region with length
scales comparable to the atomic size. Certainly the boundaries of
such regions are not fully impenetrable, and particles are then able
to escape from the region. Generally the confining surface is not
spherical. The energy spectrum of H and He-atom in
penetrable/impenetrable spherical and ellipsoidal boxes have been
studied by several authors who have employed several approaches
\cite{10,11,chemH,chemHelipse,chemHe1,chemH1,Eur.J.Phy}. In all of
these models, it is assumed that confining potential only acts up to
a cut-off boundary, and do not disturb the atomic Hamiltonian inside
the region. Under this simplification, the problem of confined
atoms, would be equivalent to the problem of many-body
Schr\"{o}dinger equation with a given boundary condition.
Practically, finding the solution of many-body Schr\"{o}dinger
equation with specified boundary condition, is not an easy task even
by numerics. As a simple but less accurate method to solve this
equation, self consistent Hartree-Fock, or variational techniques
are usually taken into account. Another methodology for solving
many-body problems, is Quantum Monte Carlo (QMC) method. This method
is among the most powerful methods in obtaining the ground state
energy of a many-body system, such as two-electron atoms constrained
in spherical impenetrable boxes \cite{QMC-JPhys}. In this article,
we have used a familiar variant of QMC technique the so-called {\it
diffusion} QMC method \cite{DMC,thez} to obtain the ground state
energy of the simplest confined atoms i.e.; H and He. Our motivation
arises from the fact that in real confined atoms, the location of
the nucleus does not exactly coincides with the centre of boxes.
Since the analytical approaches fail in these cases, employing
numerical techniques is inevitable. In order to take into account
the effects of size, shape and the potential of the boundary on the
ground state energy of confined H and He atoms, we have examined two
geometries namely spherical, and ellipsoidal compartments with
different sizes and potentials at their boundaries. It is worthwhile
to note that the problem of H-atom inside a sphere or cylinder, as a
model for H absorption in fullerene or carbon nanotube
\cite{Hcnt,neekamal,Sen}, has found applications in the condensed
matter physics. Moreover, the problem of confining He, as the
simplest problem combining repulsive electron-electron interaction
effect with the repulsion resulting from confinement, could be of
interest from theoretical point of view. For example in the finite
temperature we expect an ellipsoidal shape for C$_{60}$ molecule
instead of a spherical geometry and in this case one can model this
molecule by a prolat-shaped cage \cite{connerade}.

Our new findings can be summarized as follows.
 We have proposed a set of trial wave functions for H- atom and He -atom in a penterable/impenterable spherical
 and ellipsoidal box and computed the effect of off centrality on the ground state energy for various
 radii and penetrability. We recall that for these types of atoms, the number of electrons are
less than three. Under this condition, it would not be vital to use
the fixed node approximation as is used for fermionic systems
\cite{fixnode}. We first try to solve the centrally symmetric
problems, and then will consider generalizations which involve the
effect of displacing the nucleus off the geometrical centre of the
compartments. The paper has the following organization: In sec. II
we briefly describe our used model. Sec. III reviews the basic
ingredients of DMC method and the specific algorithm used in the
paper. In Sec. IV we have introduced the used trial wavefunctions
and sec. V includes the main results. The paper is ended in sec. VI
by some concluding remarks and discussions.

\section{The model}

Consider a simple atom subjected to a constant external potential
$V_0$ which acts beyond the exterior region of an encompassing
compartment. The time-independent Schr\"{o}dinger equation for such
H-atom is:
\begin{eqnarray}
H\psi({\bf{R}})=E\psi({\bf{R}}). \label{eqn1}
\end{eqnarray}
where
\begin{eqnarray}
H({\bf{R}})=T({\bf{R}})+V_{en}({\bf{R}})+V_{ee}({\bf{R}})+V_{ext}({\bf{R}}),
\label{eqn2}
\end{eqnarray}
here $\bf{R}=(\bf{r},\bf{x})$ and ${\bf{r}}$, $\bf{x}$ denote the
coordinates of the electrons and the nucleus respectively. $T$ is
the kinetic energy term for electrons. The nucleus is assumed
immobile due to its large mass compared to electron's mass. $V_{en}$
is the Coulombic interaction potential between the electrons and the
nucleus and $V_{ext}$ is the external applied potential. $V_{ee}$ is
the electron-electron interaction for many electron systems which is
zero beyond boundaries. The external potential is zero inside the
compartment and constant $V_0$ on its surface and beyond.

\section{Diffusion Monte Carlo method for solving Schr\"{o}dinger equation}
In this section we briefly outline the main ingredients of the
Diffusion Monte Carlo scheme. The essence of DMC lies in mapping the
Schr\"{o}dinger equation in imaginary time onto a diffusion-like
equation \cite{qmcsolid,DMC}. Through an analytic continuation of
time $t\rightarrow -i\tau$ to imaginary values, the
Schr\"{o}dinger equation will appear in the following form:\\
\begin{eqnarray}
\hbar \frac{\partial\psi({\bf R},\tau)}{\partial\tau}=
\emph{D}\nabla^2_{{\bf r}} \psi({\bf R},\tau) -V({\bf R})\psi({\bf
R},\tau). \label{eqn5}
\end{eqnarray}
In which the constant $\emph{D}$ is $\frac{\hbar}{2m}$. This
equation describes the diffusion of an ensemble of interacting
particles subject to death and/or growth processes depending on the
sign of $V(\bf R)$. The rate of death or growth i.e.; $-V(\bf R)$ is
not constant but rather space-dependent. In the long time limit,
$\psi(\bf R,\tau)$ tends to a unique value $\psi_s(\bf R)$ which is
a functional of the potential $V(\bf R)$ which is identical to the
ground state wave function of the original quantum mechanical
problem. In many situations, $\psi(\bf R)$ shows divergent to
infinity or convergent to zero types of behaviour. For examples if
$-V(\bf R)$ is positive, as in the hydrogen atom  with potential
$V(r)=\frac{-e^2}{r}$, Evidently $\psi_s(x)$ diverges because after
a sufficient time, the number of particle in the ensemble diverge
due to creation process. Now to remedy the problem, one adds a trial
constant $E_T$ into the imaginary time
Schr\"{o}dinger equation as follows:\\
\begin{eqnarray}
\hbar\frac{\partial\psi({\bf R},\tau)}{\partial\tau}=
\emph{D}\nabla^2_{{\bf r}} \psi({\bf R},\tau) +(E_T-V({\bf
R}))\psi({\bf R},\tau). \label{eqn6}
\end{eqnarray}
In the limit $\tau \rightarrow \infty$, $\psi({\bf R},\tau)$ will
only tends to a non zero and finite value provided $E_T$ coincides
with the system ground state $E_0$. To make this approach
applicable, we can exploit different methods of solving
diffusion-like equation such as random walk and Green function.
Although a significant advantage of the diffusion Monte Carlo is
that one does not need to specify a trial wave function, to speed up
the implementation of the method and to avoid errors arising in the
problems containing singularities in the potential, it would be
helpful to use a guiding trial wavefunction. More specifically, in
the positions of singularities, where $V({\bf R})$ diverges, the
creation/annhilation rate $E_T-V({\bf R})$ becomes tremendously
large and can cause undesirable fluctuation in the number of
diffusive particles which makes the algorithm unstable. To smooth
this singularities, one uses a time independent trial wavefunction
$\psi_T({\bf R})$. To implement this idea, we introduce a function
$\rho({\bf R} ,\tau)$ defined as $\rho({\bf R},\tau)=\psi_T({\bf
R})\psi({\bf R},\tau)$. Substituting $\psi({\bf R},\tau)$ in terms
of $\rho$ in equation (4) gives the following equation:

\begin{eqnarray}
\hbar\frac{\partial\rho({\bf R},\tau)}{\partial\tau}&=&\emph{D}
\nabla_{{\bf r}}[ \nabla_{{\bf r}} -F({\bf R}) ]\rho({\bf R},\tau)\nonumber\\
&-&(E_L({\bf R})- E_T)\rho({\bf R},\tau). \label{eqn7}
\end{eqnarray}

The term $F({\bf R})=\frac{2\nabla\psi_T({\bf R})}{\psi_T({\bf R})}$
is called {\it Fokker-Planck} force and corresponds to force
function which drifts the walkers away from regions where
${|\psi_T({\bf R})|}^2$ is small. The local energy function $E_L(\bf
R)$ is defined as $\frac{-\emph{D}\nabla_{{\bf r}}^2 \psi_T({\bf
R})}{\psi_T({\bf R})} +V({\bf R}).$ An appropriate choice of the
trial wavefunction $\psi_T({\bf R})$ gives rise to non singular
$E_L(\bf R)$. The method we have taken into account is based on the
Green's function approach to equation (5). We recall that the
Green's function of equation (5) can not be obtained exactly.
However, the problem can be treated using an approximate Green's
function for short time evolution. The appropriate approximated
Green's function is divided
into two terms \cite{qmcsolid} which are given below:\\

\begin{eqnarray}
G_D({\bf R},{\bf R'},\Delta \tau)=\frac{exp(-\frac{({\bf R'}-{\bf
R}- \emph{D} F({\bf R})\Delta\tau)^2}{4D~\Delta\tau})}{(4\pi
\emph{D}~\Delta\tau)^{3N/2}} +O(\Delta \tau)^2, \label{eqn8}
\end{eqnarray}

for diffusive term and

\begin{eqnarray}
G_{B}({\bf R}^{\prime}, {\bf R}, \Delta\tau)=\exp[-\Delta
\tau(\frac{E_L({\bf R})+E_L({\bf
R}^{\prime})}{2}-E_T)],~\label{eqnb1}
\end{eqnarray}
for the branching term. This is the part of the Green's function
responsible for the stochastic creation/annihilation process. To
improve the error of the order $ (\Delta\tau)^2$, we have used the
Metropolis procedure which will be explained shortly. Throughout the
paper, we have used the following atomic units $\hbar=m=e=1$. We
note that the atomic unit of length is the Bohr radius
$\frac{\hbar^2}{me^2}=0.529$ {\AA} ; the time unit is
$\frac{\hbar^3}{me^4}=2.419\times 10^{-2}$ fs and the energy unit is
$Hartree$ i.e.; $\frac{me^4}{\hbar^2}=27.25$ eV.

\subsection {DMC steps}

In what follows we give the algorithmic steps for implementation of
the method described above. Firstly, we initialize the system. A
number $N_0$ of non interacting particles (random walkers) are
distributed in space. It is preferable to distribute them according
to an approximate guess of the ground state wavefunction. Then the
position and number of walkers are updated according to the
following stochastic rules. These rules are applied synchronously to
all the walkers of the ensemble. During an MC step, each walker
execute a guided discrete time random walk. To this end, we choose a
constant time interval $\Delta\tau$ between walks. The walk length
and its direction are stochastic variables. Since the walk is
directed, it contains a directed plus a random component. More
precisely, the new
position of each particle is update according to the following rule:\\
\begin{eqnarray}
{\bf R'}={\bf R}+D\Delta\tau F({\bf R}) +\eta\sqrt{\Delta\tau}~,
\label{eqn9}
%\left(\sqrt{\kappa/\sigma_t}q_m\right)}\right]\right\
\end{eqnarray}
In which $\eta$ is a Gaussian random number with zero mean and unit
variance. The term $D\Delta\tau F({\bf R})$ corresponds to the
guided part. The attempted move is accepted if the Metropolis test
is successful. This test introduces a ratio $w$ given as below:
\begin{eqnarray}
w=\frac{G({\bf R},{\bf R'},\Delta\tau)\rho({\bf R'})}{G({\bf
R'},{\bf R},\Delta\tau)\rho({\bf R})} \simeq \frac{G({\bf R},{\bf
R'},\Delta\tau)\psi_T^2({\bf R'})}{G({\bf R'},{\bf
R},\Delta\tau)\psi_T^2({\bf R})}.
 \label{eqn10}
\end{eqnarray}
The move is accepted with the probability $w$. The next stage
simulates creation/annihilation processes. After accepting the move,
we evaluate the branching factor $q=e^{- \Delta\tau (\frac{E_L({\bf
R})+E_L({\bf R'})}{2}-E_T )}$. Let $[q]$ denote the integer part of
$q$ and $q-[q]$ its fractional part. When $q$ is greater than one,
with the probability $q-[q]$, $[q]+1$ replicas and with the
probability $1-(q-[q])$, $[q]$ replicas are replaced at the updated
location of the particle to the ensemble of particles. Note that if
$q$ is less than one, with probability $1-q$ the particle
annihilates. After performing the above diffusion/branching
processes for all the particles in the ensemble, and hence updating
the ensemble number $N$, the final MC step consists of updating the
value of $E_T$. Among various choices of updating rule, we have chosen the following one:\\
\begin{eqnarray}
E_T = <E_L> -\frac{N-N_0}{N_0~\Delta\tau } \label{eqn11}
\end{eqnarray}
Where $N_{0}$ refer to the initial number of walkers, $N$ denotes
the updated number of walkers. The average is an ensemble average.
This adjusting of $E_T$ is essential to avoid large fluctuation in
the walkers number. After a sufficient MC steps, the time series
$E_T$ reaches a steady state. The ground state of the system is
obtained by averaging $E_T$ over many MC steps.

\section{\ Trial wavefunctions }
\subsection{\ Spherical penetrable boundaries: Centered atom}

In this paper, we have used trial wavefunctions which are
appropriately chosen associated to the spherical boundary condition.
We note that the potential takes a constant value $V_0$ on the
boundary and beyond it. For the spherical boundary $a$ denotes the
radius of the sphere. The impenetrable case has already been done in
\cite{QMC-JPhys}.

When the boundaries are penetrable, for H-atom, we
have used an iterative scheme to find the appropriate trial wavefunction.
This scheme is implemented according to the following steps: at
first we employ the same unbiased method used in \cite{DMC}, for
producing the radial wavefunction for all values of $V_0$ and $a$.
After obtaining the required data for the radial wavefunction, in
the second step we find the best fit of the following function to
these data. We propose the trial wave function as follows:

\begin{eqnarray}
\Psi_{T}^H(r,V_0)=  Y_3~exp(-Y_1r^3-Y_2r^2-Y_0r). \label{eqnV}
\end{eqnarray}

The fitted function $\Psi_T^H(r)$ is in turn used as the trial
wavefunction. We note the parameters $Y_i$'s are functions of $V_0$
and $a$. By using the cusp condition, $\psi_T'(r)+\psi_T(r)=0$, we
can determine one of the above parameters such as $Y_0$ which is
equal to one. This condition guarantees singularity cancelation at
regions where $r$ is small. The values of the fitting parameters are
exhibited for some values of $a$ and $V_0$ in Table I. In this case
$E_L$ for $r<a$ turns out to be
 \begin{eqnarray}
E_{LH}(r, V_0)=\dfrac{1}{2}\left[  (6Y_1r+2Y_2)-{(3Y_1r^2+2Y_2r+Y_0)}^2 \right]\nonumber\\
-\dfrac{1}{r} \left[ -3Y_1r^2-2Y_2r-Y_0+1 \right]\nonumber
\label{eqn70}
\end{eqnarray}
For exterior regions where $r>a$ the local energy should change
to $E_{LH}(r,V_0)+\dfrac{1}{r}+V_0$. Note that the local energy is
obviously finite in the region $r>a$. In the case of He atom inside
a penetrable sphere the trial wave function is
$\Psi_{T}^H(\textbf{r}_1)\Psi_{T}^H(\textbf{r}_2)J(r_{12})$. Here
the Jastrow function has the form $J(r_{12})=
e^\frac{r_{12}}{2(1+\alpha r_{12})}$ with $r_{12}$ as the distance
between two electrons and $\alpha=0.2$. The cusp condition leads us
to choose $Y_0=2$ . Then the local energy when $r_1<a$~and~$r_2<a$
is written as below

\begin{eqnarray}
E_{LHe}(r, V_0)= E_{LH}(r_1, V_0)+E_{LH}(r_2, V_0)+\dfrac{1}{r_{12}}\nonumber\\
 -\sum_{i=1}^{2}\left[  \dfrac{\nabla_i\Psi_i}{\Psi_i}.\dfrac{\nabla_i J}{J}+\dfrac{1}{2} \dfrac{{\nabla_i}^2 J}{J}\right]\nonumber
 \label{eqn110}
\end{eqnarray}
 where as before $J=J(r_{12})$ and $\Psi_i=\Psi_{T}^H(r_i,V_0)$ according to equation (11).
  In the case of $r_1>a$~or~$r_2>a$ when one electron goes outside of the sphere the local energy is written as
 \centerline{  \\$E_{LHe}(r, V_0)+\left[\dfrac{1}{r_k}-\dfrac{1}{r_{12}}+V_0 \right]$}\\
  In the above equation $k$ refers to the index of electron in outside region.
    Finally when both electrons are outside the region, i.e.,  $r_1>a~$ and $~r_2>a$ the local energy should be written as
   \centerline{\\$E_{LHe}(r, V_0)+\left[\sum\dfrac{1}{r_i}-\dfrac{1}{r_{12}}+V_0 \right].$}

\subsection{\ Spherical boundaries: Off-centred atom}

Now we discuss the off-centrality in the spherical geometries. When
the off-centrality is small and the boundaries are impenetrable we
have obtained, for H atom, the trial wavefunction by implementing a
shift $d$ along the z-direction:
$r'=\sqrt{x^2+y^2+(z-d)^2}=\sqrt{\rho^2+(z-d)^2}$.

\begin{eqnarray}
\Psi_{TD}^H(\textbf{r}')= e^{(-r'+\lambda(z,d) r'^2)} j_0(k r),
\end{eqnarray}
where $j_0(kr)$ is the zero order Bessel-function with $k=\pi/a$.
The first term plays the role of off-centrality and the second term
insures the vanishing of the wave function on the boarders. Defining $f=e^{-r'+\lambda(z,d) r'^2}$
and using cylindrical coordinates, the local energy is found to have the following form:

 \begin{eqnarray}
E_{LD}(\rho,z)=
-\dfrac{1}{2}\left[\dfrac{d^2f/d\rho^2}{f}+\dfrac{1}{\rho}
\dfrac{d\ln f}{d{\rho}} \right] -\dfrac{d^2f/dz^2}{2f}
+\nonumber\\\dfrac{k^2}{2}- \dfrac{\hat r}{f}\left[
\hat\rho\dfrac{df}{d\rho}+\hat{z}\dfrac{df}{dz}\right].\left[k~cot(kr)-\dfrac{1}{r}
\right],
 \label{eqn15}
\end{eqnarray}
where $\hat r=\hat\rho+\hat z$.  For keeping the local energy finite
every where especially at $r=a$ one needs to solve an ordinary
differential equation for $\lambda(z,d)$. Such solution is a complex
function which in the range $d\leq 0.5a$ has a simple linear
behaviour, $\lambda=Az+B$ where $A$ and $B$ depend on $d$. Table II
includes the values of $A$ and $B$ for many geometries.

When the boundaries are penetrable and off-centrality is small, we
have used $\Psi_{T}^H(r',V_0)$ with the same functional form as in
the equation (11). The corresponding parameters $Y_i$'s are taken
from Table I.

In the case of small off-centred He atom within an impenetrable
spherical box we have used the following trial function:

\begin{eqnarray}
\Psi_{TD}^H(\textbf{r}'_1)\Psi_{TD}^H(\textbf{r}'_2)J(r_{12}),
\end{eqnarray}
where ${r_i}'=\sqrt{x_i^2+y_i^2+(z_i-d)^2}$ and $r_{12}$ is defined as the distance between two electron as before.

\subsection{\ Ellipsoidal boundaries}

In the case of ellipsoidal geometries first we recall the equation
 of a standard ellipsoid which has been prolated in the z-direction by defining the function
$R=\sqrt{\dfrac{x^2+y^2}{a^2}+\dfrac{z^2}{b^2}}$. When $R=1$ we have an ellipsoid surface
  with the semi-major axis $a$ and the semi-minor axis $b$. Without loosing generality we assume $b>a$.
For small eccentricity i.e, $\epsilon=\sqrt{\dfrac{b^2}{a^2}-1}$,
our trial wave function for H atom within an impenetrable ellipsoidal boundary
is
\begin{eqnarray}
\Psi_{Te}^H(r,R)=e^{(- r+\beta r^2)} j_0(k R).
\end{eqnarray}
 $k$ should be equal to $\pi$ while
$\beta=[{2\sqrt{\dfrac{2a^2+b^2}{3}}}]^{-1}$. These values have been
determined by keeping the local energy finite and reaching to the
same function of spherical case when $a=b$. For H atom in penetrable
boundaries we have used $\Psi_{T}^H(R,V_0)$ using the functionality
analogous to equation (11).

In the case of He atom in a prolated impenetrable ellipsoid with
small eccentricity we have used the following wave
 function $\Psi_{Te}^H(r_1,R_1)\Psi_{Te}^H(r_2,R_2)J(r_{12})$.
For He atom in penetrable ellipsoidal boundaries one can replace the
firs two terms of equation (14) by $\Psi_{T}^H(R_1,V_0)$ and
$\Psi_{T}^H(R_1,V_0)$ respectively.

\section {Results}

Prior to presenting our results, it would be illustrative to discuss
the numerical errors in the DMC method. Principally, there are two
types of errors which limit the accuracy of most DMC calculations:
(a) statistical or sampling errors associated with the limited
number of independent sample energies used in determining the ground
state energy and (b) the systematic errors associated with finite
time-step $\Delta \tau$, round-off in computing, imperfectness of
random number generators etc. For a spacial boundary condition like
the spherical boxes with penetrable boundaries, we know the energies
value (steady state energy) from the analytical calculations
(\cite{QMC-JPhys}), therefore, minimizing the variance of energy,
can give us a range of time step. We have used the minimum of these
time step, i.e., $\Delta \tau$=0.005. The number of walkers was
taken $N_0=5000$ in our simulation.

\subsection{\ Spherical box }

The first quantity that we have calculated is the electronic ground state
wavefunction (see Fig. 1). Due to spherical symmetry, the solution
of the Schr\"{o}dinger equation can be solved by separating the
variables so that the wavefunction is represented by the product
$\psi (r,\theta ,\phi )=\Phi (r)\Theta (\theta )\varphi (\phi )$.
Furthermore, because the ground state wavefunction of centred H-atom
in a spherical geometry has radial symmetry, one can replace
$\psi $ with $r\Phi(r)$ for simplicity. For an atom confined in the
centre of an impenetrable and very small sphere ($V_{0}=\infty $ and
a=1), the wavefunction dramatically differs from that of free atom. Accordingly, the maximum of the wavefunction
shifts towards smaller $r$, which also causes larger slope in its
curvature (see Fig. 1(a) ). On the boundary, the wavefunction tends
smoothly to zero. Since kinetic energy corresponds to the Laplacian
of the wavefunction, this enhancement of the slope consequently
leads to the enhancement of the ground state energy. The situation
is significantly different for large encompassing regions where the
radius of the sphere is notably greater than one atomic unit. As can
be inferred from Fig. 1(b), where a=4.0, the electronic wavefunction
deviates slightly from the electronic wavefunction of a free H-atom.
In this case the wavefunction tends to zero with small slope.
Moreover, the influence of $V_0$ has been suppressed.

An important feature of the electronic wavefunction in small
compartments (when $a=1$) is its dependence on the penetrability
$V_0$. By decreasing the potential $V_0$, the
tendency of an electron for being near the boundary increases and
for pure penetrable sphere ($V_{0}=0.0$) the probability of
finding the electron near the boundary becomes maximum. In Fig. 2(a)
we have plotted the ground state energy versus the strength of
penetrability for several radii. Apparently, the ground state energy
strongly depends on the penetrability when the size of bounded
region is comparable to the volume of a free H-atom, i.e., a$\leq$ 1.
Nonetheless, for larger regions and larger radii, ground state
energy is almost independent of $V_{0}$.

To improve our understanding, in the  set of coluored two
dimensional figures, we have depicted the behaviour of electronic
distribution (see Fig. 3-5) for some values of $V_0$ at $a=1$ in the
$x-z$ plane. As can be seen, by increasing $V_0$, the electronic
distribution becomes more localized around the nucleus and keeps its
symmetry independent of $V_0$. In the case of fully
penetrable boundary, electron has access to exterior regions. The
blue regions represent a smaller probability for finding the
electron, whereas the red regions represent higher probabilities.
The electron does not penetrate to the exterior regions in deeper
confining potential barrier.

\subsection{Off-centred atoms}

One of the most significant advantages of DMC method lies in its
simplicity in being applied to various boundary conditions. This
property allows us to investigate the effect of displacement of
H-atom along the $z$-direction on the ground state energy of the
system. Our motivation in considering this geometrical case is
two-fold: first, in real systems (such as confined atoms in a
metallic bubble) one can not fix the atoms on the geometrical centre
and second, due to mathematical difficulties arising from the
analytical treatments. Figure 2(b) illustrates the results of this
investigation for three values of $V_{0}$. The figure depicts the
ground state energy of a H-atom confined in a sphere with radius
$a=1$ versus the nucleus displacement $d$ (in atomic unit) from
centre of the sphere. Generically, the off-centre effects become
more considerable when the displacement $d$ exceeds nearly half of
the radius $a$. Table III contains our results for ground state
energy. Our energy results for the case $d=0$ and impenetrable
boundaries are in good agreement with those in \cite{QMC-JPhys}. For
the case of penetrable boundaries our results are in well agreement
with the analytical results in \cite{chemH}.
For a deeper understanding, we exhibited the dependence of
the energy on the displacement $d$ in Fig. 2(c) for impenetrable sphere.
For larger $a$ the deviation of the energy occurs in larger displacement $d$.

For off centred helium atom all the results are shown in table IV. The
general behaviour is qualitatively analogous to the H- atom.
However, due to electron-electron interaction, geometrical effects
are suppressed compared to H- atom case. Our results for $d=0$ and
impenetrable boundaries are in reasonable agreement with the results
given in \cite{QMC-JPhys}. It has been verified that if one uses a
local and negative penetrability in the boundaries when the atom moves
towards the boarder, the electron can cross the well depending on
the well deep \cite{neekamal}.

\subsection{\ Ellipsoidal box }

Let us now investigate the ground state energy of a H- and a He- atom
confined within an ellipsoidal compartment. We consider a symmetric
type of ellipsoid characterized by parameters $a$ and $b$ which are
semi-major and semi-minor axes respectively. For simplicity, we have used small
eccentricity, i.e, $b=1.1 a$ and $b=1.5 a$ for several $a$ and
$V_0$. We recall that nucleus is placed in the centre.

The ground state energy of H-atom confined in larger ellipsoidal
cavity (large $a$ and $b$) tends to the energy of the ground state
of free H-atom, -0.5 , see table V.  When we increase $V_0$ to a
higher value we observe that energy increases especially for
smaller $a$. For a fixed $a$ , larger $b$ has larger energy in all
cases. For an ellipsoid with small semi-major axis, the ground state
energy will be more affected by the surrounding potential.

For many-particle systems, finding an exact solution of the
Schr\"{o}dinger equation is merely possible and approximative
methods should inevitably be taken into account. The ground state
wavefunction of He-atom is node-less and the Pauli exclusion principle does not
play a significant role. For this reason the above simple DMC method is
reasonable. In table VI we have reported energy values of He-atom in
several ellipsoidal geometries. All used parameters such as $a$, $b$
 are defined similar to the H-atom case. By increasing $a$
and $b$ the ground state energy asymptotically tends to the lowest
state energy of free He-atom i.e. $-2.903$ au. In small radius the
energies have larger values, because electrons are closer to each
other and to the nucleus. For higher values of $V_0$, two electrons
are confined only inside the box and interact via Columbic forces,
therefore the energy should be increased. The elimination of the
electron-electron interaction decreases the ground state energy in
comparison to the case where this interaction is on.

In the absence of nucleus, systems with the larger values of $a$ and
$b$ behave as a 2-dimensional confined electron gas which are vastly
studied in quantum dots \cite{quantumdot}.

\section{\ Summary}

In this paper we have evaluated the ground state energies and
wavefunctions of confined Hydrogen and Helium atoms by the method of
Diffusion Monte Carlo. In particular, we have studied spherical and
ellipsoidal confining compartments. Ground state values crucially
depend on the size of confining compartments. In larger
compartments, the results asymptotically approach to the values of
free atoms. The effect of displacing the nucleus from the
geometrical centre of the confining region has also been
investigated. This effect is of theoretical as well as practical
interest. We have shown that the ground state energies for both H
and He atoms inside the ellipsoidal box significantly depend on the
position of nucleus and the penetrability of boundaries. The merit of
DMC is that it simply can treat an arbitrary shape of boundaries.
Our future work will be devoted to find a reasonable model for
confined atoms within a carbon nano-tube and correspondingly
applying QMC methods on such a systems with a cylindrical
confinement. Work along this line is in progress and will be
reported elsewhere.

\newpage
%\begin{figure*}
%\begin{center}
%\includegraphics[scale=0.6]{fig1.epsi}
%\caption{\footnotesize (Color online) (a) H-atom inside a
%penetrable/impentrable sphere. Dashed lines show the penetration
%boundary. (b) Variation of energy versus the radius of encompassing
%compartment for three values of $V_0$. Both DMC and analytical
%results(\cite{chemH}) have been shown in figure. }
%\end{center}
%\end{figure*}

\newpage
\begin{table*}
\begin{tabular}{|c|cccc|cccc|cccc|cccc|cccc|cccc|}
\hline
$a$&$V_0$ &$Y_1$  & $Y_2$ & $Y_3$&$V_0$&$Y_1$  & $Y_2$ & $Y_3$&$V_0$&$Y_1$&$Y_2$& $Y_3$&$V_0$&$Y_1$ & $Y_2$ & $Y_3$&$V_0$&$Y_1$ & $Y_2$ & $Y_3$&$V_0$&$Y_1$& $Y_2$ & $Y_3$    \\
\hline\hline
 1.0&&0.0&0.01&2.0&&0.0&0.22&3.19&&0.08&0.93&4.99&& 0.01&0.9&4.68&&0.06&0.14&3.73&&0.02&0.95&4.93 \\
 2.0&&0.0&0.04&2.23&&0.0&0.24&2.89&&0.01&0.47&3.41&&0.01&0.52&3.35&&0.01&0.62&3.31&&0.08&0.54&3.39\\
3.0&0.0&0.0&0.02&2.07&1.0&0.03&0.0&2.14&2.0&0.01&0.3&2.64&3.0&0.01&0.24&2.57&4.0&0.09&0.51&2.89&5.0&0.01&0.83&3.31\\
4.0&&0.0&0.42&2.61&&0.01&0.64&2.87&&0.01&0.31&2.54&&0.01&0.16&2.23&&0.03&0.61&2.75&&0.02&0.23&2.50\\
5.0&&0.01&0.44&2.59&&0.01&0.74&2.90&&0.01&0.5&2.66&&0.01&0.48&2.51&&0.01&0.11&2.24&&0.01&0.41&2.61\\
 \hline
\end{tabular}
\caption{ Values of the fitted parameters needed for trial
wavefunctions in spherical penetrable box.} \label{ResultList}
\end{table*}
\newpage
\begin{table*}
\begin{tabular}{|ccc|ccc|ccc|ccc|ccc|}
\hline
 &a=1  &  &   &a=2 &  & &a=3  &   &  &a=4 & &   &a=5 & \\
\hline
d &A  & B &  d  & A& B & d &A  & B  & d  & A& B& d  & A& B\\
\hline
0.1 &0.035 & 0.498& 0.2 & 0.009& 0.249& 0.3 & 0.004& 0.166&  0.4 & 0.002 & 0.124& 0.5 & 0.001  & 0.099 \\
0.2 &0.072 & 0.490&0.4 & 0.018& 0.245& 0.6 & 0.008& 0.163&  0.8 & 0.005 & 0.123 &1.0 & 0.003  & 0.098 \\
0.3& 0.106 &0.478& 0.6 & 0.026& 0.239& 0.9 & 0.012& 0.160& 1.2 & 0.007 & 0.119 &1.5 & 0.004  & 0.096\\
0.4& 0.134 &0.464 & 0.8 &0.033& 0.232&1.2 & 0.015& 0.155&1.6 & 0.008  & 0.116&2.0 & 0.005  & 0.093\\
0.5& 0.152& 0.447 &1.0 &0.038& 0.224& 1.5 & 0.017& 0.149& 2.0 & 0.010 & 0.112 &2.5 & 0.006 & 0.089 \\

\hline
\end{tabular}
\caption{ Values of coefficients A and B in sec. IV.B} \label{ResultList}
\end{table*}
\newpage

\begin{figure*}
\begin{center}
\includegraphics[scale=0.6]{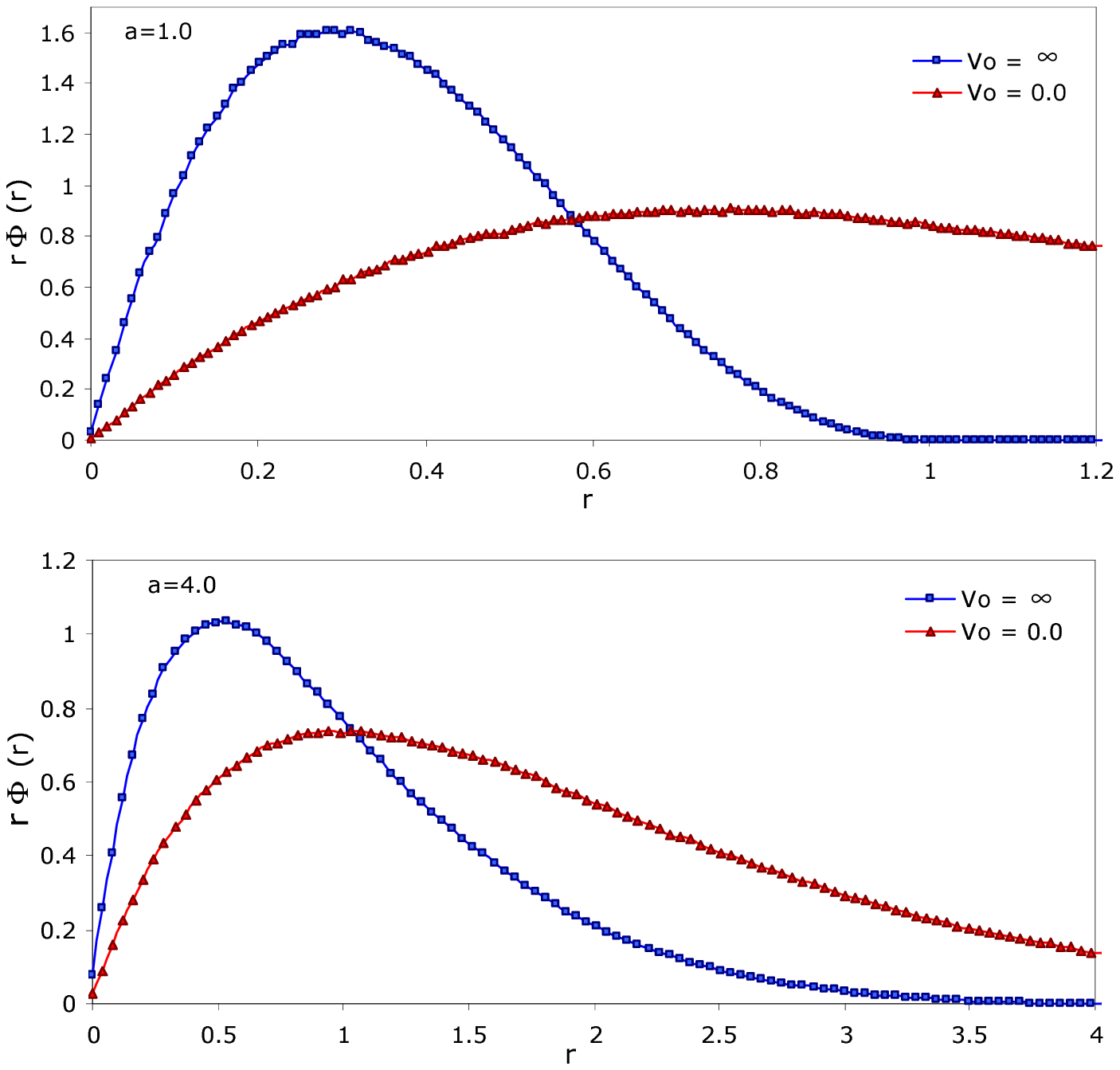}
\caption{\footnotesize (Color online) The electronic ground state
wavefunction of H-atom confined in a sphere for two values of $V_0$.
$a=1$ and $4.0$ in (a) and (b) respectively. In both figures $d=0$.}
\end{center}
\end{figure*}
\newpage

\begin{figure*}
\begin{center}
\includegraphics[scale=0.6]{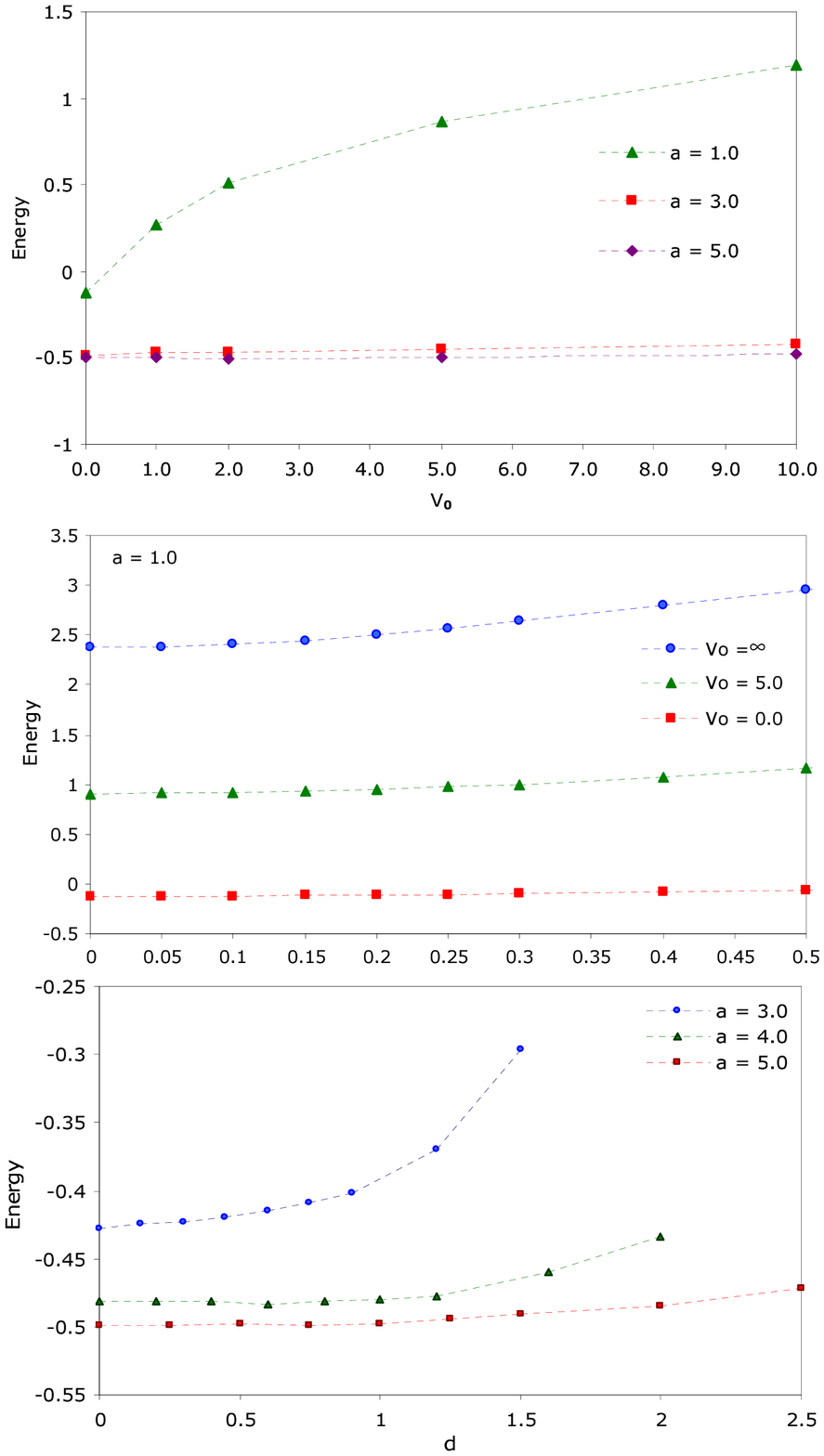}
\caption{\footnotesize (Color online) (a) Ground state energy of H-atom versus penetrability in an spherical compartment for various radii when d = 0.
(b) Ground state energy of confined H-atom for different penetrability coefficients versus off-centre distances d when a = 1.0. (c) Ground state energy of a H-atom confined in an impenetrable spherical compartment for various values of a versus off-centre distance d for V0 =$\infty$.}
\end{center}
\end{figure*}

\newpage
\begin{figure*}
\begin{center}
\includegraphics[scale=0.55]{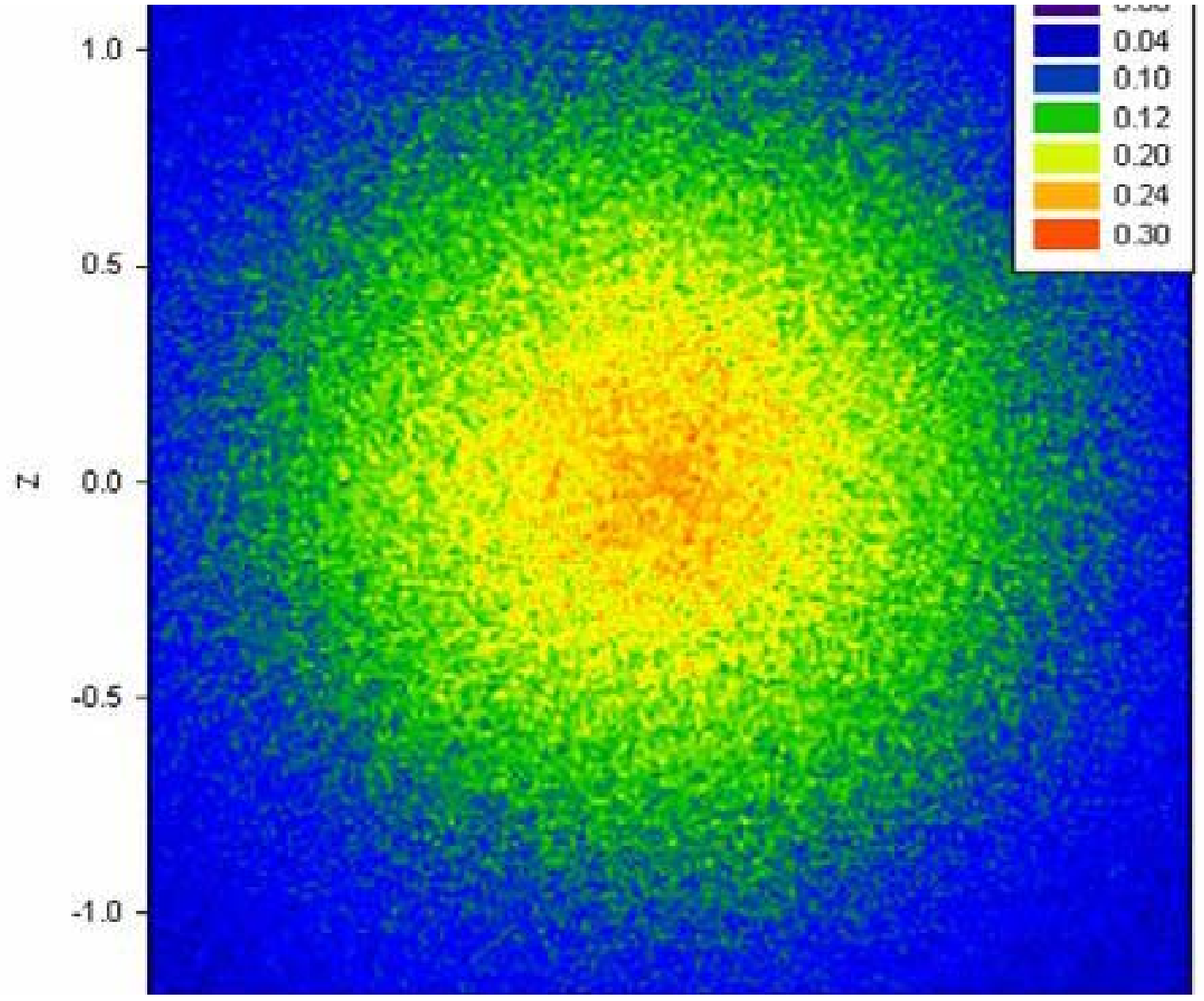}
\caption{\footnotesize (Color online) Electronic distribution of
H-atom inside a spherical compartment for $V_0=0$ and
fixed $a=1$ when $d=0$.}
\end{center}
\end{figure*}
\begin{figure*}
\begin{center}
\includegraphics[scale=0.4]{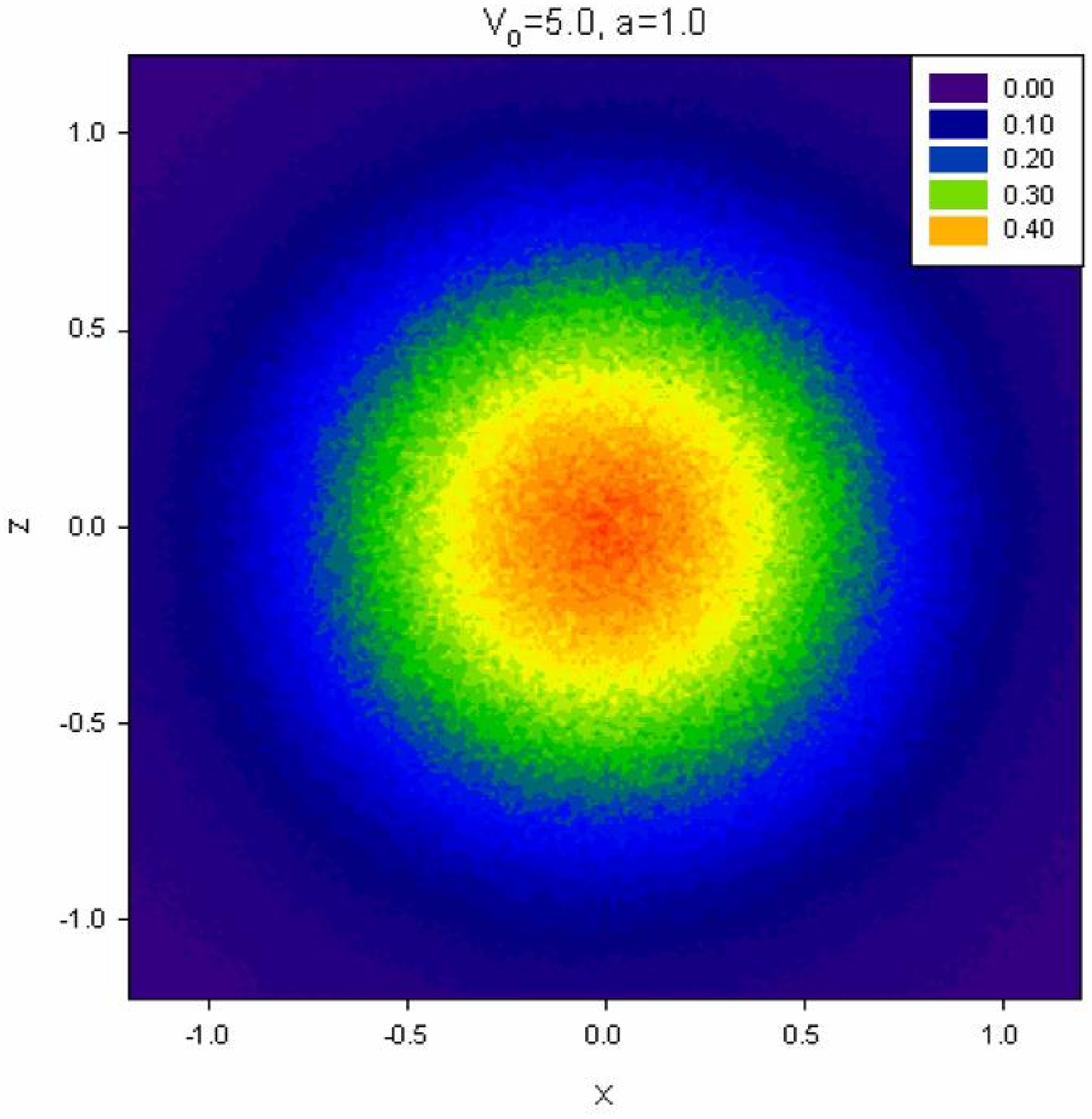}
\caption{\footnotesize (Color online) Electronic distribution of
H-atom inside a spherical compartment for $V_0=5$ and
fixed $a=1$ when $d=0$.}
\end{center}
\end{figure*}
\begin{figure*}
\begin{center}
\includegraphics[scale=0.55]{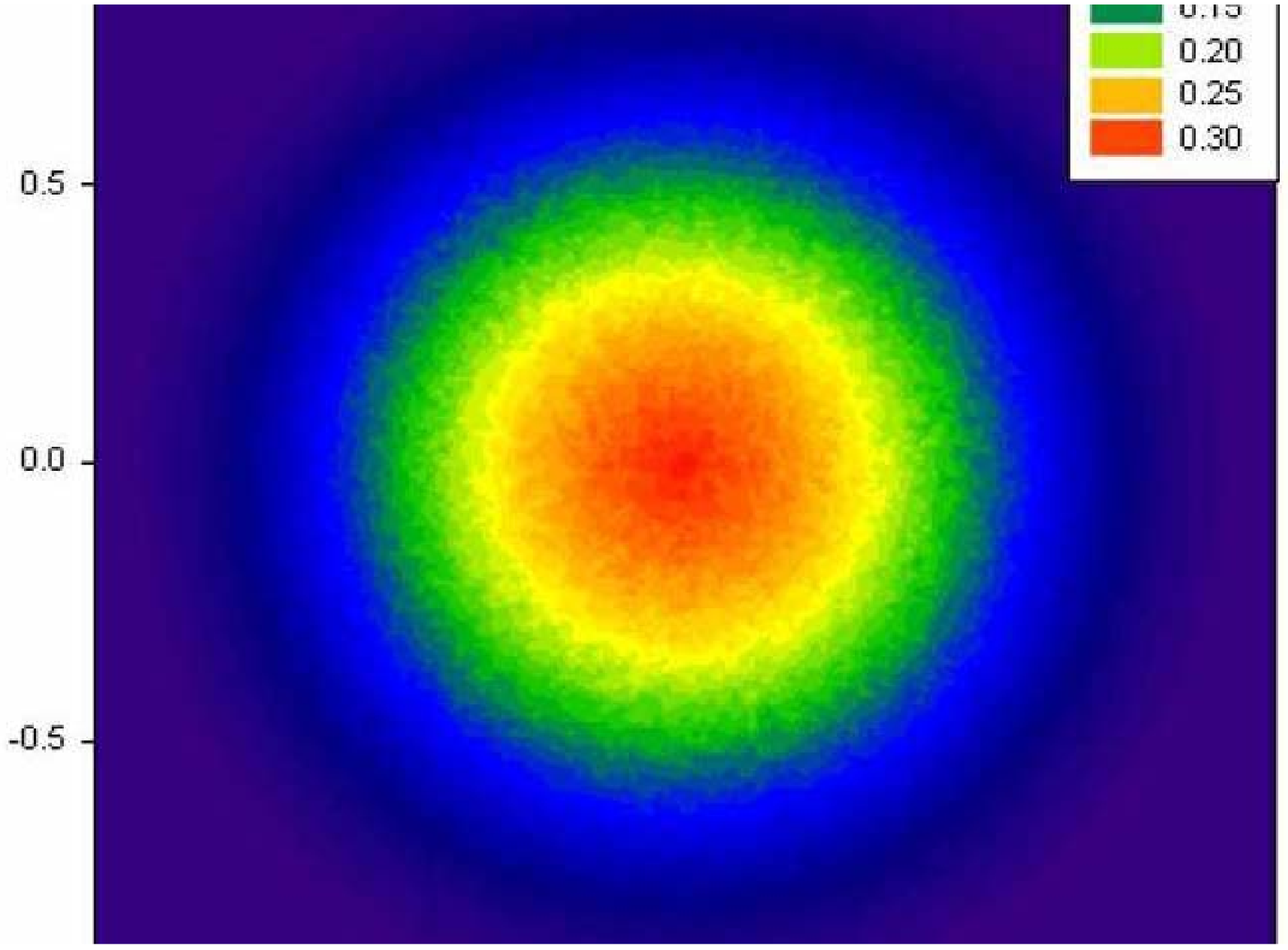}
\caption{\footnotesize (Color online) Electronic distribution of
H-atom inside a spherical compartment for $V_0=/infty$ and
fixed $a=1$ when $d=0$.}
\end{center}
\end{figure*}

\begin{table*}
\begin{center}
\begin{tabular}{|c|c|c|c|c|}
\hline
 $a$ &  $d$ & $V_0=0.0$ &  $V_0=5.0$&  $V_0= \infty$ \\
\hline
1.00&  0.00  &  -0.1250(7) &  0.822(7)  &   2.1760(6)\\
    &  0.10  &  -0.122(8) &  0.878(8)  &   2.5070(7)\\
    &  0.20  &  -0.116(9) &  0.918(4)  &   2.5970(8)\\
    &  0.30  &  -0.102(5) &  0.974(6)  &   2.6210(6)\\
    &  0.40  &  -0.082(9) &  1.040(6)  &   2.711(5)\\
    &  0.50  &  -0.059(7) &  1.1450(6)  &   2.863(1)\\
\hline
2.00&  0.00  &  -0.430(3) &  -0.182(1) &  -0.1250(6)\\
    &  0.20  &  -0.4280(7) &  -0.163(3) &  -0.027(8)\\
    &  0.40  &  -0.414(2) &  -0.1500(8) &  -0.043(7)\\
    &  0.60  &  -0.399(5) &  -0.113(6) &  -0.0260(3)\\
    &  0.80  &  -0.394(2) &  -0.028(1) &   0.021(1)\\
    &  1.00  &  -0.383(2) &   0.0600(8) &   0.110(2)\\
\hline
3.00&  0.00  &  -0.489(6) &  -0.4510(6) &  -0.425(7)\\
    &  0.30  &  -0.4840(6) &  -0.446(4) &  -0.399(8)\\
    &  0.60  &  -0.479(8) &  -0.4370(5) &  -0.385(6)\\
    &  0.90  &  -0.467(4) &  -0.428(4) &  -0.384(3)\\
    &  1.20  &  -0.463(8) &  -0.404(4) &  -0.371(3)\\
    &  1.50  &  -0.4551(4) &  -0.351(9) &  -0.296(5)\\
\hline
4.00&  0.00  &  -0.496(3) &  -0.489(3) &  -0.481(9)\\
    &  0.40  &  -0.493(7) &  -0.475(9) &  -0.470(3)\\
    &  0.80  &  -0.491(8) &  -0.449(1) &  -0.465(2)\\
    &  1.20  &  -0.489(9) &  -0.470(7) &  -0.459(4)\\
    &  1.60  &  -0.480(7) &  -0.408(1) &  -0.463(5)\\
    &  2.00  &  -0.470(4) &  -0.391(4) &  -0.435(4)\\
\hline
5.00&  0.00  &  -0.499(9) &  -0.498(6) &  -0.492(2)\\
    &  0.50  &  -0.494(7) &  -0.490(2) &  -0.488(1)\\
    &  1.00  &  -0.485(2) &  -0.479(6) &  -0.494(7)\\
    &  1.50  &  -0.491(8) &  -0.478(8) &  -0.486(4)\\
    &  2.00  &  -0.490(6) &  -0.469(4) &  -0.490(8)\\
    &  2.50  &  -0.499(7) &  -0.471(5) &  -0.471(6)\\
\hline

\end{tabular}
\end{center}
\caption{The ground state energy of confined $H$-atom for various values of
penetrability coefficients and off-center distances $d$.} \label{ResultList}
\end{table*}

\newpage

\begin{table*}
\begin{center}
\begin{tabular}{|c|c|c|c|c|}
\hline
 $a$ &  $d$ & $V_0=0.0$ &  $V_0=5.0$&  $V_0= \infty$ \\
\hline

1.00&  0.00  &  -2.2290(7) &  -0.910(5) &   1.012(4)\\
    &  0.10  &  -2.228(8) &  -0.949(6) &   1.105(6)\\
    &  0.20  &  -2.269(2) &  -0.965(2) &   1.175(9)\\
    &  0.30  &  -2.241(8) &  -0.921(4) &   1.477(5)\\
    &  0.40  &  -2.250(8) &  -0.941(6) &   1.926(2)\\
    &  0.50  &  -2.247(4) &  -0.916(8) &   2.602(7)\\
\hline
2.00&  0.00  &  -2.757(5) &  -2.652(6) &  -2.603(6)\\
    &  0.20  &  -2.774(7) &  -2.619(9) &  -2.496(3)\\
    &  0.40  &  -2.747(7) &  -2.615(2) &  -2.448(7)\\
    &  0.60  &  -2.723(2) &  -2.6300(8) &  -2.382(6)\\
    &  0.80  &  -2.759(7) &  -2.642(8) &  -2.247(1)\\
    &  1.00  &  -2.781(3) &  -2.622(9) &  -2.008(8)\\
\hline
3.00&  0.00  &  -2.831(4) &  -2.776(1) &  -2.871(8)\\
    &  0.30  &  -2.797(8) &  -2.791(6) &  -2.754(6)\\
    &  0.60  &  -2.801(1) &  -2.805(7) &  -2.749(4)\\
    &  0.90  &  -2.781(9) &  -2.7820(3) &  -2.735(4)\\
    &  1.20  &  -2.780(2) &  -2.757(2) &  -2.703(1)\\
    &  1.50  &  -2.809(7) &  -2.806(6) &  -2.607(2)\\
\hline
4.00&  0.00  &  -2.816(5) &  -2.803(1) &  -2.892(6)\\
    &  0.40  &  -2.771(6) &  -2.804(9) &  -2.800(1)\\
    &  0.80  &  -2.806(8) &  -2.830(1) &  -2.795(8)\\
    &  1.20  &  -2.7980(8) &  -2.787(4) &  -2.781(8)\\
    &  1.60  &  -2.814(1) &  -2.792(9) &  -2.778(1)\\
    &  2.00  &  -2.801(4) &  -2.789(9) &  -2.7380(4)\\
\hline
5.00&  0.00  &  -2.812(8) &  -2.798(2) &  -2.902(7)\\
    &  0.50  &  -2.815(1) &  -2.806(1) &  -2.8710(8)\\
    &  1.00  &  -2.8280(7) &  -2.776(2) &  -2.820(9)\\
    &  1.50  &  -2.765(4) &  -2.7750(5) &  -2.797(5)\\
    &  2.00  &  -2.797(9) &  -2.793(7) &  -2.812(8)\\
    &  2.50  &  -2.821(6) &  -2.791(9) &  -2.800(4)\\
\hline

\hline
\end{tabular}
\end{center}
\caption{The ground state energy of confined $He$-atom for various
values of penetrability coefficients and off-centre distances $d$.}
\label{ResultList}
\end{table*}

\begin{table*}
\begin{center}
\begin{tabular}{|c|c|c|c|c|c|c|c|c|}
\hline
 $a$ &  $b$ & $V_0=0.0$ &  $V_0=5.0$&  $V_0= \infty$&  $b$ & $V_0=0.0$ &  $V_0=5.0$&  $V_0= \infty$ \\
\hline

1.00 & 1.10 & -0.1320(2)&  0.819(6) &  1.832(1) & 1.50 & -0.179(2) &  0.605(9) &  1.460(7)\\
1.50 & 1.65 & -0.352(1)& -0.013(1) &  0.269(8) & 2.25 & -0.367(9) & -0.095(9) &  0.133(8)\\
2.00 & 2.20 & -0.428(6)& -0.288(8) & -0.184(3) & 3.00 & -0.443(7) & -0.338(4) & -0.257(8)\\
2.50 & 2.75 & -0.466(7)& -0.415(1) & -0.374(1) & 3.75 & -0.476(9) & -0.428(7) & -0.397(2)\\
3.00 & 3.30 & -0.482(1)& -0.450(5) & -0.434(3) & 4.50 & -0.485(4) & -0.466(1) & -0.452(2)\\
3.50 & 3.85 & -0.490(6)& -0.473(4) & -0.477(4) & 5.25 & -0.489(3) & -0.480(2) & -0.481(2)\\
4.00 & 4.40 & -0.490(6)& -0.484(5) & -0.492(3) & 6.00 & -0.492(3) & -0.492(3) & -0.494(6)\\
4.50 & 4.95 & -0.491(3)& -0.490(6) & -0.489(2) & 6.75 & -0.497(2) & -0.493(1) & -0.4940(6)\\
5.00 & 5.50 & -0.490(9)& -0.490(2) & -0.493(7) & 7.50 & -0.486(5) & -0.497(8) & -0.500(8)\\

\hline
\end{tabular}
\end{center}
\caption{The ground state energy of H-atom confined in an
 ellipsoidal box, with $b=1.1a$ and $b=1.5a$.} \label{ResultList}
\end{table*}

\begin{table*}
\begin{center}
\begin{tabular}{|c|c|c|c|c|c|c|c|c|}
\hline
 $a$ &  $b$ & $V_0=0.0$ &  $V_0=5.0$&  $V_0= \infty$&  $b$ & $V_0=0.0$ &  $V_0=5.0$&  $V_0= \infty$ \\
\hline

1.00 & 1.10 & -1.6890(3) & -1.101(7) &  0.203(4) & 1.50 & -1.923(3) & -1.429(3) & -0.371(2)\\
1.50 & 1.65 & -2.544(6) & -2.382(3) & -2.115(6) & 2.25 & -2.611(9) & -2.502(7) & -2.274(3)\\
2.00 & 2.20 & -2.795(2) & -2.750(4) & -2.644(3) & 3.00 & -2.780(6) & -2.772(8) & -2.712(8)\\
2.50 & 2.75 & -2.881(4) & -2.8350(9) & -2.813(3) & 3.75 & -2.851(9) & -2.851(6) & -2.820(7)\\
3.00 & 3.30 & -2.859(7) & -2.874(5) & -2.880(3) & 4.50 & -2.904(3) & -2.869(3) & -2.8780(6)\\
3.50 & 3.85 & -2.868(3) & -2.888(7) & -2.908(2) & 5.25 & -2.885(2) & -2.924(8) & -2.866(8)\\
4.00 & 4.40 & -2.877(8) & -2.894(4) & -2.865(7) & 6.00 & -2.8900(6) & -2.881(8) & -2.885(9)\\
4.50 & 4.95 & -2.8480(1) & -2.895(3) & -2.889(2) & 6.75 & -2.883(9) & -2.864(9) & -2.860(5)\\
5.00 & 5.50 & -2.887(2) & -2.9150(8) & -2.899(2) & 7.50 & -2.889(4) & -2.870(3) & -2.869(3)\\

\hline
\end{tabular}
\end{center}
\caption{The ground state energy of He-atom confined in an
ellipsoidal box, with $b=1.1a$ and $b=1.5a$.} \label{ResultList}
\end{table*}

\end{document}